\documentstyle[aps,floats,epsfig]{revtex}
\begin{document}

\newcommand{\bce}{\begin{center}}
\newcommand{\ece}{\end{center}}
\newcommand{\boldF}{\mbox{\boldmath $F$}}
\newcommand{\boldV}{\mbox{\boldmath $V$}}
\newcommand{\boldpi}{\mbox{\boldmath $\pi$}}
\def\lsim{\mathrel{\rlap{\lower4pt\hbox{\hskip1pt$\sim$}}
    \raise1pt\hbox{$<$}}}       
\def\gsim{\mathrel{\rlap{\lower4pt\hbox{\hskip1pt$\sim$}}
    \raise1pt\hbox{$>$}}}

\title{Scalar-isoscalar excitation in dense quark matter}

\author{Prashanth Jaikumar$^1$ and Ismail Zahed$^2$} 

\address{$^1$Physics Department, McGill University, Montreal QC H3A
	 2T8, Canada\\
	 $^2$Department of Physics and Astronomy, SUNY at Stony Brook, 
	 New York 11794-3800}

\date{\today} \maketitle

\begin{abstract}
We study the spectrum of scalar-isoscalar excitations in the
color-flavor locked phase of dense quark
matter. The sigma meson in this phase appears as a four-quark state
(of diquark and anti-diquark) with a well-defined
mass and extremely small width, as a consequence of it's small coupling
to two pions. The quark particle/hole degrees of freedom also
contribute significantly to the correlator just above the threshold
2$\Delta$, where $\Delta$ is the superconducting gap.      
\end{abstract}
%-------------------------------------------------------------------------
\section{Introduction}

\label{sec_intro}
%-------------------------------------------------------------------------
The generally accepted theory of strong interactions, Quantum Chromodynamics
(QCD), has several remarkable and unique features, one of them being
the spontaneous breakdown of chiral symmetry ($\chi_{SB}$) in the
vacuum, which dictates the physics of the low energy hadronic world
(light pseudoscalars). This is borne out by the predictive power of
current algebra techniques, chiral effective Lagrangians and chiral
perturbation theory to pion mediated processes in the threshold
region~\cite{Adler68,Weinberg79,Gasser84}. More recently, a consistent
calculational scheme extending to the resonance region has been
developed~\cite{Yam96}. Chiral symmetry breaking in the vacuum is
characterized by the existence of an order parameter, the quark
condensate $\langle: q\bar{q} :\rangle$. Long wavelength axial-like phase
fluctuations of the condensate correspond to the familiar pions, while
the quantum fluctuations of it's norm represent the sigma meson
$\sigma\sim \langle:(q\bar{q})^2:\rangle$.  Empirically, the $\sigma$
meson is seen as a broad resonance in the $I=J=0$ channel of $s-$ wave
$\pi\pi$ scattering, and it's large width (600-1000 MeV), comparable
to it's mass (400-1200 MeV)~\cite{ParticleDB2000} follows from it's
strong coupling to two pions. From a theoretical standpoint, the
$\sigma$ meson is invoked in any linear realization of chiral
symmetry breaking such as in the Gell-Mann Levy linear sigma
model~\cite{GellMann60} or the Nambu Jona-Lasinio
model~\cite{NJL61}. Despite the difficulties in the experimental resolution of
the sigma meson, there are strong theoretical arguments for it's
importance in vacuum hadronic physics~\cite{K01}.
It's spectral function is expected to change when finite temperature
and/or baryon density effects are included~\cite{Hatsuda99,Schuck00},
these effects being linked to the evolution of the chiral condensate
with temperature ($T$) and baryon density ($\mu_B$). Qualitatively
similar effects can be reproduced with $p-$wave renormalization of
the two pions to which the 'bare sigma' of mass $\sim 800$ MeV
couples~\cite{Schuck88}. A particular and consistent feature of these
studies, corroborated by experimental data from the CHAOS experiment
on $A(\pi,2\pi)$ knock-out reactions, is the spectral enhancement at
the $2m_{\pi}$ threshold, indicative of a softening of the $\sigma$
meson and perhaps a partial restoration of chiral symmetry at finite
baryon density~\cite{Hatsuda99}. In this way, the sigma meson is more
likely to be identified clearly in hot and dense matter.
\vskip 0.1cm Thus far, the spectral enhancement has been investigated
at temperatures and densities around the chiral phase transition. At
much higher densities corresponding to perhaps (5-10)$\varrho_0$ or
more, ($\varrho_0$ being the saturation density of nuclear matter) the
physics of the $\sigma$-meson is as yet unexplored, though certainly
accessible in light of recent advances in theories of high density
quark matter, which have shown that a BCS-like quark phase is
energetically preferred. The idea of color superconducting quark
matter is now a few decades old~\cite{Frau78,Barrois77,Bailin84} but
renewed interest in this field began only a few years
ago~\cite{Alford98,Rapp98,alf,rapp,Schafer99,Evans99}. Since single
quarks are expected to pair with large gaps of a hundred MeV or
more~\cite{Alford98,Rapp98}, it is the low energy excitations of the
system that determine the response to small external
perturbations. For the case of pairing among two light flavors, chiral
symmetry remains unbroken whereas for three light flavors, a
color-flavor locked (CFL) phase is favored which breaks chiral
symmetry because left- and right-handed flavor rotations are locked to
the color gauge field. In this letter, we study the response of
CFL matter to an external scalar probe. We find a light $\sigma$ meson
($\sim 250$ MeV for large quark gaps) that is sharply peaked owing to
it's small coupling to a two-pion correlated state. Above a threshold
energy of $2\Delta$, the quark particle/hole degrees of freedom can
also couple to the external probe with appreciable strength. We
consider the possibility of dilepton and neutrino production from the
scalar mode in dense matter. These processes constitute the leptonic
decay of the sigma meson in this dense phase.
\section{Scalar correlator in the CFL phase I: Weak coupling}
\begin{figure}[!h]
\vspace{0.5cm}
\bce
\epsfig{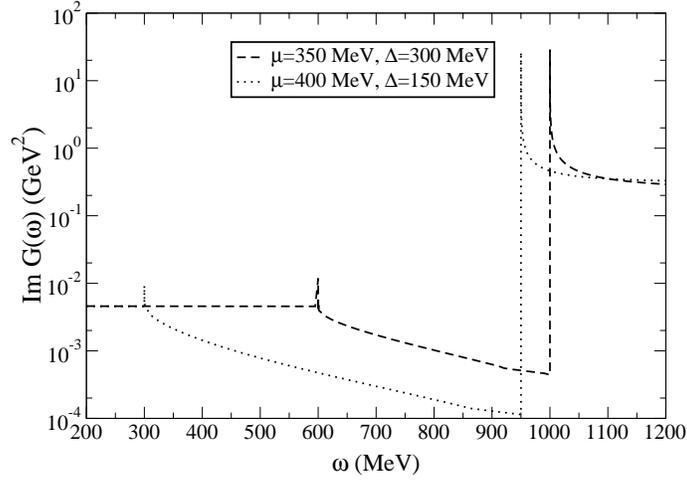}
\ece
\vspace{0.3cm}
\caption{Cut of the loop involving
the quarks and diquarks. The response has the usual
$\bar{q}q$ threshold as well as a pair-breaking threshold at $\omega=2\Delta$.}
\label{figps1}
\end{figure}

The scalar probe will couple to a $q\bar{q}$ excitation. In this
section, we therefore evaluate the contribution of the quark loop to
the scalar correlator
$\langle T\{\bar{\psi}\psi(x)\bar{\psi}\psi(y)\}\rangle_{CFL}$.
In momentum space, the quark loop is given by
\begin{equation}
G(Q)=-\frac{1}{2}{\rm Tr}_{q,s,c,f,NG}\biggl[\Gamma~S(K)~\Gamma~S(P)\biggl] 
\end{equation} 
Here, $Q=(Q_0,{\bf Q})$ is the external momentum, while
$K=q+Q/2, P=q-Q/2$ with $q$ being the internal momentum of the loop. The trace is performed over the internal momentum $q$, spin $s$,
color $c$, flavor $f$ and Nambu-Gorkov $NG$ indices. The scalar vertex
is simply $\Gamma=\biggl(\begin{array}{cr} {\bf 1} & 0 \\ 0 &
-{\bf 1}
\end{array}\biggr)$ where ${\bf 1}={\bf 1}_c\times{\bf 1}_f\times{\bf
1}_s$. The identity matrix in flavor reads ${\rm diag}(1,1,0)$ so that
only $u,d$ quarkfields are introduced. This is in accordance with the
light quark nature of the sigma meson at zero density. At high
density, the s-quark can however, appear
inside the loop in between two gap insertions on the same quark
line. The structure of the entries of the propagation matrix $S$ in
the Nambu-Gorkov basis are detailed in the Appendix. As shown there,
in the large $\mu$ limit, simplified expressions may be used for the
quark propagator.  As we are interested in generating a spectral
function, we will study the energy dependence of the correlator and
work at zero three-momentum $Q=(Q_0,{\bf 0})$.  The result of tracing
over the internal symmetries yields (see Appendix {\bf A} for details)
\begin{eqnarray}
G(Q_0)&=&\int\frac{d^4q}{(2\pi)^4}(T_1+T_2+T_3+T_4); \\
T_1&=&\frac{-1}{2}{\rm Tr}_{s,c,f}\biggl[{\bf 1}S_{11}(K){\bf
1}S_{11}(P)\biggr] \label{qloop}\nonumber\\
&=&\biggl\{-6\frac{(k_0+k_{||})(p_0-\mu-|{\bf
p}|)}{(k_0^2-\epsilon_k^2)(p_0^2-\bar{\epsilon}_p^2)}-\frac{3(m_u^2+m_d^2)}{4\mu^2}\biggl[\biggl(\frac{p_0+p_{||}}{p_0^2-\epsilon_p^2}\biggr)^2+\frac{(k_0+k_{||})(p_0+p_{||})}{(k_0^2-\epsilon_k^2)(p_0^2-\epsilon_p^2)}\biggr]\nonumber\\
&+&\frac{m_u^2+m_d^2+2m_s^2}{\mu^2}\biggl(\frac{|G(K)|}{k_0^2-\epsilon_k^2}\biggr)^2\biggr\}
+ (K\leftrightarrow P)\quad.\\
T_3&=&\frac{1}{2}{\rm Tr}_{s,c,f}\biggl[{\bf 1}S_{12}(K){\bf
1}S_{21}(P)\biggr]\nonumber\\
&=&\frac{2m_um_d}{\mu^2}\biggl(\frac{G(K)G^*(P)}{(k_0^2-\epsilon_k^2)(p_0^2-\epsilon_p^2)}\biggr)\quad.
\end{eqnarray}   
$T_2$ and $T_4$ follow from $T_1$ and $T_3$ respectively under the
substitutions $\pm\mu\leftrightarrow\mp\mu,\pm|{\bf q}|\leftrightarrow
\mp |{\bf q}|, \pm G(q)\leftrightarrow \mp G^*(q)$. The terms in $T_1$
which come from the diagonal elements of the Nambu-Gorkov propagator
have the following intepretation. The first term is the
particle(/hole)-antiparticle(/antihole) contribution arising from the
massless part of the quasiparticle propagator. The particle-hole
contribution from a massless propagator is finite only for non-zero
${\bf Q}$, therefore it does not appear here. The reason is that
$\bar{\psi}\psi$ flips chirality whereas at ${\bf Q}=0$, the particle
and hole have the same helicity (same as chirality for massless
particles). The second term owes it's origin to mass insertions that
can flip chirality (two mass insertions on only one of the two lines
of the loop gives the first term in the square bracket while one mass
insertion each in both lines gives the second). For quarks near the
Fermi surface, the gap affects the propagation. The third term
includes the possibility of having two gap insertions in between two
mass insertions on the same quark line.  $T_3$ originates from the
off-diagonal parts of the Nambu-Gorkov propagator. As such, it
involves gap insertions on each quark line as well as mass insertions
to flip chirality. The delineation of these terms allows us to
understand the associated contribution to the spectral function that
relates to the imaginary part of the correlator. The imaginary parts
are obtained after analytic continuation as $Q_0=\omega+i\epsilon$,
and are plotted in Fig.(\ref{figps1})
for 2 sets of quark chemical potentials and gaps.    
\vskip 0.2cm We begin by considering the first two terms of $T_1$ and
$T_2$ which represent the quark-antiquark response from the 'bulk' of
the matter. Assuming the antiquarks to be ungapped (since the
anti-gap dependent pieces in the propagator appear only at ${\cal
O}(\mu_q^{-2})$), these terms are relevant to energies greater than
$2\mu+\Delta$ (an antiquark needs energy $\approx 2\mu$ to be freed
while the quark needs only an energy $\Delta$ (see
Eqn.(\ref{qqbar})).\\ The terms inside the square bracket from $T_1$
and $T_2$ which come from mass insertions (see Eqn.(\ref{massins}))
can be reliably estimated only for typical energies greater than the
light quark mass because of the small mass expansion of the quark
propagators. This contribution arises at low energies and extends up
until $\omega<2\epsilon_q$. The remaining term from $T_1$ and $T_2$
involves two gap insertions in addition to two mass insertions on the
same quark line. The resulting contribution to the spectral function
(see Eqn.(\ref{massgapins})) shows that the strange quark mass appears
at this level even though the external current involves only light
quarks . The reason is the color-flavor locked structure of the
gap. Similar caveats as for the first two terms of $T_1,T_2$ apply
here as well ($m_{u,d}<\omega<2\epsilon_q$).
\vskip 0.2cm
The contribution from the off-diagonal components of the quark
propagator involve diquarks (Eqn.(\ref{gapins})). It is proportional to
$\Delta^2$ as well as the light quark masses and
contributes significantly just above the threshold to break a pair,
i.e., $2\Delta$. As we are interested in moderately large
chemical potentials, we will rely on instanton-based non-perturbative
estimates of the gap, which may be as large as a few hundred MeV at
intermediate densities~\cite{rapp}.
\section{scalar correlator in the cfl phase II: sigma meson}
\begin{figure}[!h]
\vspace{0.5cm}
\bce
\epsfig{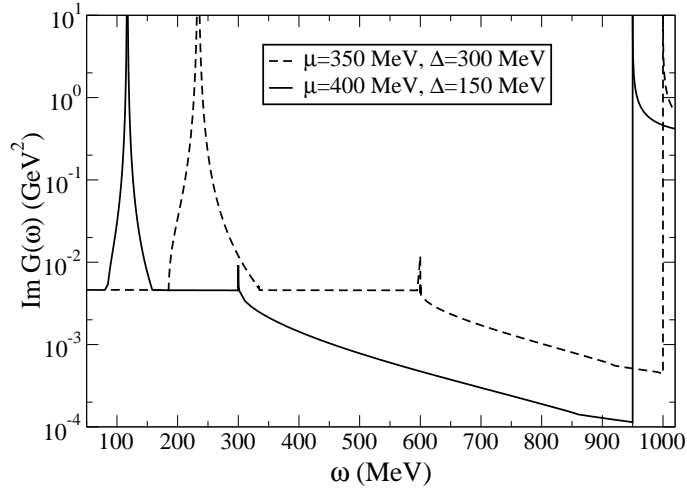}
\ece
\vspace{0.3cm}
\caption{Imaginary parts from the quark loop and the $\sigma$ meson
contribution. The latter appears as a
low mass sharp resonance.}
\label{figps3}
\end{figure}

\begin{figure}[!h]
\vspace{0.5cm}
\bce
\epsfig{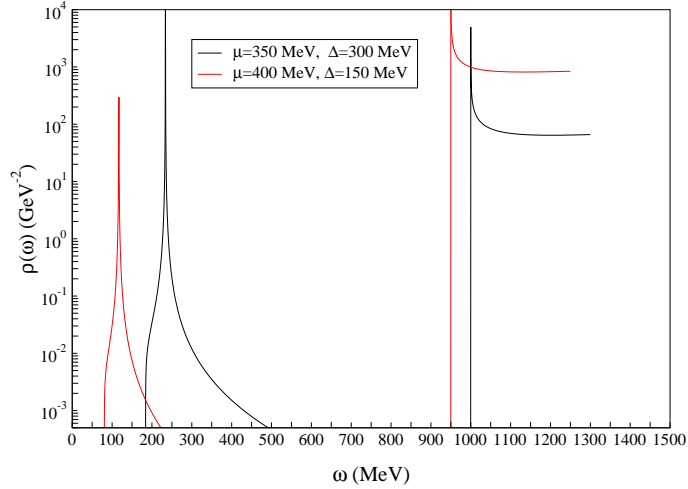}
\ece
\vspace{0.3cm}
\caption{The full spectral function including contributions from the
quark loop and the sigma meson.}
\label{figps4}
\end{figure}

Based on weak coupling analyses to leading logarithm accuracy,
effective Lagrangian approaches to QCD in the CFL phase have proven to
be useful with applications to color-flavor anomalies, pesudoscalar
meson masses, hidden local symmetry and vector meson
properties~\cite{Hong99,Casal99,Son,Schafer,Jai}. Here, we use the
effective Lagrangian to obtain the coupling of the $\sigma$ meson in
the CFL phase to a scalar probe, and we model its width in the dense
phase via a Breit-Wigner resonance through its coupling to two
generalized pions~\cite{Hong99}. In this way, we show that the $\sigma$ meson
appears as a sharp low mass excitation and makes an important
contribution to the scalar correlator.
\vskip 0.2cm The low energy excitations of the CFL phase are two
singlet modes associated with $U(1)_B$ and $U(1)_A$ symmetry breaking
(corresponding to vector and axial baryon number fluctuations
respectively) and an octet of Goldstone modes associated with chiral
symmetry breaking. The octet is described by a low energy theory that
bears strong resemblance to chiral perturbation
theory~\cite{Casal99,Son}. The Goldstone modes are parametrized by a
$3\times3$ unitary matrix $U$ which is a color singlet, transforming
under $SU(3)_L\times SU(3)_R$ as $U\rightarrow g_LUg_R^{\dag}$, where
$U$ is related to axial-like fluctuations of the left and right handed
diquark fields:
\begin{equation}
 L^{ai}\sim\epsilon^{abc}\epsilon^{ijk}\langle q_L^{bj}q_L^{ck} 
 \rangle^{*},\quad R^{ai}\sim\epsilon^{abc}\epsilon^{ijk} 
 \langle q_R^{bj}q_R^{ck}\rangle^{*},\quad {\rm and} \quad U=LR^{\dag}\quad.
\end{equation}
The fields $L$ and $R$ carry color $(ijk)$ and flavor $(abc)$, 
and transform under $g_f \subset SU(3)_f ~{\rm and}~ g_c\subset SU(3)_C$, 
(where $f$ denotes left or right handed flavor) as
\begin{equation}
L\rightarrow g_LLg_C^{\dag}\,\quad {\rm and} 
\quad R\rightarrow g_RRg_C^{\dag}\,,
\end{equation}
respectively. The low energy effective theory is governed 
by the Lagrangian~\cite{Schafer}
\begin{eqnarray}
{\cal L}_{eff}&=&\frac{f_{\pi}^2}{4}{\rm Tr}\biggl[\nabla_0
U\nabla_0 U^{\dag} - v_{\pi}^2\partial_i U \partial_i
U^{\dag}\biggr]\nonumber \\
&+&\biggl[A_1{\rm Tr}(MU^\dag){\rm Tr}(MU^\dag)+A_2{\rm
Tr}(MU^\dag MU^\dag)+A_3{\rm Tr}(MU^\dag){\rm Tr}(M^\dag U)+{\it
H.C.}\biggr]+... \label{leff}\quad, 
\end{eqnarray}
where $\nabla_0$ includes Bedaque-Sch\"afer effective mass terms in
addition to the usual partial time derivative, and $v_{\pi}^2$ denotes
the Goldstone boson velocities.\footnote{In the weak coupling limit,
we have $v_\pi^2=1/3$~\cite{Son99}. As a consequence of the breaking
of Lorentz invariance in matter, there are two pion decay constants:
the temporal decay constant $f_T =f_\pi$ and the spatial decay
constant $f_S=v_{\pi}^2 f_\pi$. Indeed, axial vector current
conservation demands that $f_Tw^2-f_S{\bf q}^2=0$ for an on-shell
pion.}  In~\cite{Schafer,Son99}, the coefficients $A_{1,2,3}$ in the
chiral theory have been matched to an effective high density theory at
the Fermi surface, obtained by integrating out the high momentum
modes~\cite{Hong100,Hong200}, such that the energy shifts from the
mass terms in the two theories are in agreement. Their values
are~\cite{Schafer,Son99}
\begin{equation}
A_1=-A_2=\frac{3\Delta^2}{4\pi^2},\quad A_3=0.
\end{equation}
As we are interested in the mass of the $\sigma$ meson and it's
coupling to two pions, we show that they can be obtained from the mass
terms in the following way. Although Eqn.(\ref{leff}) denotes a
non-linear sigma model, we note that the sigma field represents
fluctuations about the finite expectation value of
$\langle :(q\bar{q})^2:\rangle$. Therefore, fluctuations about the mass
terms, which appear multiplied to the chiral field $U=L^\dag R$, the
phase field of $\langle qq\bar{q}\bar{q}\rangle$, can be understood as the
sigma meson. This is analogous to fluctuating radially about the non-trivial
minimum of the potential for an $O(4)$ linear sigma model to obtain
the sigma field. The leading mass terms in the chiral effective theory
have a different form than those in vacuum chiral perturbation theory
if instanton induced interactions are suppressed at high
density~\cite{Rapp98}, leading to a restoration of $U(1)_A$
symmetry. In that case, terms like ${\rm Tr}MU$ are forbidden. The
mass terms appearing in Eqn.(\ref{leff}) instead respect the $Z_2$
symmetry of the theory. The restoration of $U(1)_A$ symmetry is also
the reason that we do not concern ourselves here with the physics of
the $\eta^\prime$ meson (This may be included in the effective
Lagrangian by considering the overall $U(1)$ phases of the condensate
as in~\cite{Schafer,Son99}). Thus, the parts of the effective
Lagrangian that are relevant to the mass and coupling (to two pions) of
the sigma are the terms proportional to $A_1$ and $A_2$. These terms
contain the coefficients of the fields $\sigma^2$ and
$\sigma\pi\pi$. Expanding the chiral field $U$ to ${O}(\pi^2)$ via the
parametrization $U=\exp (2i\pi^a\tau^a/f_\pi)$, (where the $SU(3)$
generators $\lambda^a=2\tau^a$ are normalized as ${\rm tr}[\lambda^a
\lambda^b]=2\delta^{ab}$), we find that (see Appendix {\bf C} for
details)
\begin{equation}
m_\sigma^2=\frac{6\Delta^2}{\pi^2},\quad
g_{\sigma\pi\pi}=\frac{m_\sigma^2(\delta m)}{f_\pi^2} \label{sigmaeff}
\end{equation} 
where $\delta m =m_d-m_u > 0$. The non-linear sigma model is obtained
by integrating out the sigma field, therefore, no kinetic term for it
appears, which prevents from properly normalizing the $\sigma$ mass
term. However, we have in mind a linear realization of chiral symmetry
breaking in order to assess the contribution of the $\sigma$ meson to
the scalar correlation function. Thus, the chiral field can as well be written
as $U=\sigma+i\gamma_5\pi.\tau$, leading to the canonical form of the
kinetic term $\frac{1}{2}(\partial_{\mu}\sigma)^2$. The spectral
function of the $\sigma$ meson is now modelled as a Breit-Wigner
resonance with a mass $m_\sigma$ and a width determined by it's
coupling to a two-pion state, as detailed in Appendix {\bf C} and
displayed in Fig.(\ref{figps2}).
\begin{figure}[!h]
\vspace{0.5cm}
\bce
\epsfig{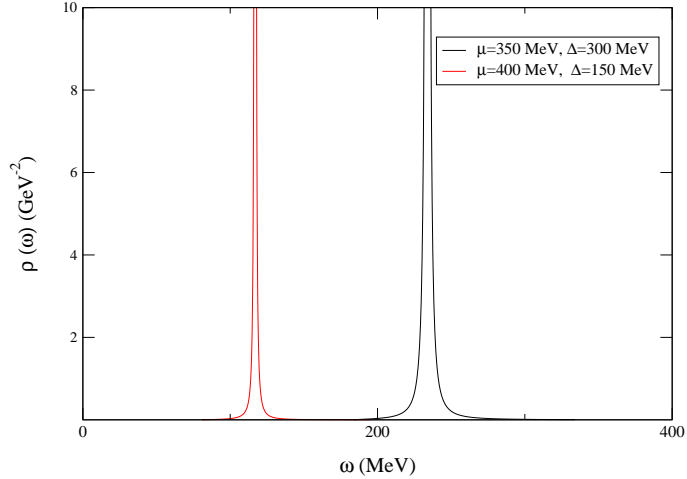}
\ece
\vspace{0.3cm}
\caption{Spectral function of the $\sigma$ meson as modelled by a
Breit-Wigner resonance, with mass and width determined from the chiral
effective Lagrangian Eqn.(\ref{leff}).}
\label{figps2}
\end{figure} 
\vskip 0.2cm
In order to assess the contribution of the $\sigma$ meson to the
scalar correlator, we need to compute the overlap of $\bar{\psi}\psi$
with the $\sigma$ field. This is not straightforward since the
$\sigma$ involves diquark fields. The procedure involves an extension
of the method discussed in Ref.~\cite{Gatto01} to couple fermion
fields to the Goldstone modes, which also involve diquark fields. For the
CFL realization, the invariant coupling of fermion fields to the
Goldstones can be written as~\cite{Gatto01}
\begin{equation}
-\frac{\Delta}{2}\sum_{I=1,..,3}{\rm Tr}[(\psi
L^\dag)^TC\epsilon_I(\psi
L^\dag)\epsilon_I]=-\frac{\Delta}{2}\sum_{I,I^\prime=1,..,3}{\rm Tr}[\psi^TC\epsilon_IL_{II^\prime}\psi\epsilon_{I^\prime}]\quad,
\end{equation}
where $\psi$ is to be considered as a $3\times3$ matrix in color and
flavor space, and $(\epsilon_I)_{ab}=\epsilon_{Iab}$ with $a,b$ denoting
color indices. In a model where chiral symmetry breaking is realized linearly, the $\sigma$ meson is introduced as
the norm of the chiral field, with the pions being the fluctuations
along the directions that are degenerate with respect to axial
rotations. In that case, the chiral field
$U=\sigma+i\gamma_5\pi.\tau$. Since $L$ can be expanded as
$L=1+\frac{i\pi.\tau}{2f_\pi}+...$ and $U=L^\dag R=L^2$ (for a unitary
choice of gauge), an analogous expression for the
fermion coupling to $\sigma$ follows
\begin{equation}
-\frac{\gamma_5\Delta}{2f_\pi}\sum_{I,I^\prime=1,..,3}{\rm Tr}[\psi^TC\epsilon_II_2\psi\epsilon_{I^\prime}]\quad,
\end{equation} 
where $I_2={\rm diag}(1,1,0)$ since the $\sigma$ meson involves the
$u,d$ flavors. Thus, only $I,I^\prime=1,2$ will give a
finite result for the coupling. We can now proceed to evaluate the
overlap between the coherent $\sigma$ excitation and the scalar
operator $\bar{\psi}\psi$.
\begin{equation}
J_{\sigma\bar{\psi}\psi}=\frac{\Delta}{2f_\pi}\sum_{I,I^\prime=1,2}{\rm
Tr}[{\bf 1}~S(P)~T~S(K)] ;\quad T=\biggl(\begin{array}{cr} {\bf 0} &
\gamma^0\epsilon_I(I_2)_{II^\prime}\epsilon_{I^\prime}\gamma_5\gamma^0
\\  \epsilon_I(I_2)_{II^\prime}\epsilon_{I^\prime}\gamma_5 &
{\bf 0}
\end{array}\biggr)\quad.
\end{equation}
The traces are performed over Nambu-Gorkov indices and all other
internal symmetries. Particulars of the loop momentum integration 
are explained in Appendix {\bf D} and we quote the final result here
(ignoring the energy dependence of the gap).
\begin{eqnarray}
J_{\sigma\bar{\psi}\psi}&=&\frac{\Delta^2(m_u+m_d+2m_s)}{\pi^2f_\pi}\biggl[\Theta(2\Delta-\omega)2x_0+\Theta(\omega-2\Delta)
\biggl(\int_{-\mu}^{{\rm max}(-\mu,-\sqrt{\omega^2/4-\Delta^2})}+\int_{\sqrt{\omega^2/4-\Delta^2}}^{\Lambda_*}\biggr)\frac{dq_{||}}{\epsilon_q}\biggr]\quad,
\nonumber \\
x_0&=&{\rm ln}\biggl(\frac{\Lambda_*}{\Delta}\biggr) \quad,
\end{eqnarray}
with $\Lambda_*=\biggl(\frac{4\Lambda_{\perp}^6}{\pi m_E^5}\biggr)$
denoting an upper cutoff on the $dq_{||}$ integration that is set by
the leading logarithm estimate of the gap with perturbatively screened
gluons~\cite{Park99}. The $\sigma$ meson resonance contribution to
the imaginary part of the scalar correlation function is
given by
$J_{\sigma\bar{\psi}\psi}\rho_{\sigma}J_{\sigma\bar{\psi}\psi}(\omega)$,
which is shown included with the quark loop contribution in
Fig(\ref{figps3}).
The spectral function derived from the full correlator is shown in
Fig.(\ref{figps4}).
\vskip 0.2cm 
\section{$\sigma$ coupling to electroweak probes}
The relevance of scalar correlations to dilepton emission
in a dense quark medium has been addressed previously
in~\cite{Schulze}, where it was shown that resonant interactions in the
$\bar{q}q$ scalar channel yield an emission rate that is much lower
than even the perturbative $\bar{q}q$ Born rate by 2-3 orders of
magnitude. Therein, the chosen densities and temperatures were
centered in the vicinity of the chiral phase transition, and were
motivated by fits to experimental hadron abundancies measured in
ultrarelatvistic S and Pb beam collisions at SPS energies. The
dilepton rate from the vector-isovector ($\rho$) excitation in the
ordered CFL phase, which may occur at somewhat higher densities and
much lower temperatures than considered in~\cite{Schulze} was assessed
through the Hidden Local Symmetry approach in~\cite{Jai} and compared
to standard hadronic approaches with in-medium effects. Here, we can
comment on the possibility of dilepton emission from the scalar
excitation along the lines of~\cite{Schulze}. The CFL quarks couple to
the in-medium photon $\tilde{A}_{\mu}$ with strength
$\tilde{e}=e\,{\rm cos}\, \theta_{CFL}$~\footnote{${\rm
tan}\,\theta_{CFL}= 2e/\sqrt{3}g$, and $e$, $g$ denote the gauge
couplings of $A_\mu$, $G_\mu$ respectively.} via the vertex
\begin{equation}
(\Gamma^{\mu})^{ij}_{ab}=(\tilde{Q})^{ij}_{ab}\gamma^{\mu}=
\frac{1}{\sqrt{3}} \biggl[(\lambda_{8})^{ij}\delta_{ab}
-\delta^{ij}(\lambda_{8})_{ab}\biggr]
\gamma^{\mu} \quad,
\label{tildep}
\end{equation}
since $\tilde{Q}=Q\otimes 1+1\otimes Y$ where $Y$ represents the color
hypercharge operator. In eq.(\ref{tildep}), $\{i,j\}$ and $\{a,b\}$
denote flavor- and color-indices, respectively. The color-flavor trace
involved in the coupling of the $\sigma$ to the photon
$\tilde{\gamma}$ is then
Tr[$\epsilon_II_2\epsilon_{I^\prime}M\tilde{Q}$]=Tr[$\epsilon_II_2\epsilon_{I^\prime}\tilde{Q}M^{\dag}$]=0
since the sum of the $\tilde{Q}$ charge carried in the quark loop is
zero. Thus, the $\sigma-\tilde{\gamma}$ effective coupling vanishes
and the scalar mode does not contribute to dilepton emission. However,
neutrino pair emission via coupling to the $Z_0$ neutral electroweak
gauge boson can occur since the color-flavor traces are now finite,
with $\tilde{Q}$ replaced by ${\bf 1}_c\otimes (T_3(f_{L,R})-Q(f){\rm
sin}^2~\theta_W)$ where $T_3(f_{L,R})$ denotes the isospin projection
for the relevant flavor $f$ of quark with left ($L$) or right
($R$)-handedness, and $Q$ is the corresponding (normal) electric
charge. The above weak vertex is for free quarks, therefore, the
neutrino emissivity resulting from the weak decay would be penalized
exponentially as exp(-$\Delta/T$) due to the breaking of the gap.  In
reality, in the CFL phase, the weak gauge bosons ($W$ and $Z$) also
mix with the gluons through the diquark condensate which carries
electroweak charge as well as color, leading to a $\tilde{Z_0}$
massive eigenstate for the neutral gauge boson and a modified
$Z$-coupling to the quarks in the condensate, as has been discussed
in~\cite{Sannino01} for the 2SC phase. Even in this case, there would
be a suppression as exp(-$m_{\sigma}/T$) from the thermal occupation
factor of the sigma mesons in the initial momentum state. Given the
large value of $m_{\sigma}$ for a color superconducting phase in a
neutron star and the fact that the temperature $T$ of the inner (and
more dense) regions of the star rapidly drops to the keV regime within
a few minutes of it's birth, neutrino emission from this process is
not expected to be compete with processes involving weak decays of the
pseudoscalar Goldstone bosons~\cite{JPS02}. Therefore, despite the
spectral enhancement at high densities, the $\sigma$ meson in this
phase remains elusive as far as the coupling to neutral currents is
concerned.
\vskip 0.2cm
\section*{ACKNOWLEDGEMENTS}
\noindent This work was supported by the US-DOE grant DE-FG0288ER40388. PJ
acknowledges support from the Natural Science and Engineering Research
Council of Canada.
%\begin{center}
%{\bf Appendix A}
%\end{center}    
\appendix
\section{Nambu-Gorkov propagator}
In the CFL phase, all quarks acquire a gap. It
is convenient to describe their propagation in the Nambu-Gorkov
formalism by the following matrix 
\begin{equation}
S=-i\langle\mbox{\boldmath$\psi\bar{\psi}$}\rangle=\biggl(\begin{array}{cr} S_{11} & S_{12} \\ S_{21} &
S_{22}
\end{array}\biggr)
\end{equation}
in terms of the two-component Nambu-Gorkov field
\mbox{\boldmath$\psi$}=$(\psi,\psi_C)$, where $\psi$ refers to quarks and
$\psi_C=C\bar{\psi}^T$ to charge-conjugated quarks,
respectively. According to Ref.~\cite{Pisarski99}, the
general entries of the matrix read
\begin{eqnarray}
S_{11}(q)&=&-i\langle\psi(q)\bar{\psi}(q)\rangle=\biggl\{\biggl(G_0^+(q)\biggr)^{-1}-\gamma^0\Delta^{\dag}(q)\gamma^0G_0^-(q)\Delta(q)\biggr\}^{-1}
\nonumber \\
S_{12}(q)&=&-i\langle\psi(q)\bar{\psi}_C(q)\rangle=-G_0^+(q)\gamma^0\Delta^{\dag}(q)\gamma^0S_{22}(q)\nonumber
\\
S_{21}(q)&=&-i\langle\psi_C(q)\bar{\psi}(q)\rangle=-G_0^-(q)\Delta(q)S_{11}(q)\nonumber
\\
S_{22}(q)&=&-i\langle\psi_C(q)\bar{\psi}_C(q)\rangle=\biggl\{\biggl(G_0^-(q)\biggr)^{-1}-\Delta(q)G_0^+(q)\gamma^0\Delta^{\dag}(q)\gamma^0\biggr\}^{-1}
\end{eqnarray} 
with
$(G_0^\pm(q))^{-1}=q{\hskip-1.6mm}/\pm\mu_q\gamma_{0}-m$, where
$\mu_q$ denotes the quark chemical potential.
In general $m={\rm diag}(m_u,m_d,m_s)$. The spin and color-flavor($c,f$)
structure of the superconducting gap $\Delta$ is made evident as
\begin{equation}
\Delta(q)=\mbox{\boldmath$M$}G(q)\Lambda^+(q)+\mbox{\boldmath$M$}\bar{G}(q)\Lambda^-(q)
\end{equation}
where $\Lambda^\pm(q)=\frac{1}{2}(1+\mbox{\boldmath$\alpha.\hat{q}$})$
are the energy projectors and $\mbox{\boldmath$M$}=\epsilon_f^a\epsilon_c^a\gamma_5=\mbox
{\boldmath$M^{\dag}$}$ with
$(\epsilon^a)^{bc}=\epsilon^{abc}$. $\bar{G}(q)$ denotes the antiparticle-gap.\\
Using the approximations
${\bf M}^{\dag}{\bf M}\approx {\bf 1}_{cf},
q_{||}\approx|\bf{q}|-\mu$, the quasiparticle energies read
$\epsilon_q^2=q_{||}^2+|G(q)|^2,
\bar{\epsilon}_q^2=(|{\bf q}|+\mu)^2+|{\bar{G}}(q)|^2$ and the entries in
the propagator simplify as follows
\begin{eqnarray}
S_{11}(q)&=&\gamma^0\frac{q_0+q_{||}}{q_0^2-\epsilon_q^2}\Lambda^-(q)+\gamma^0\frac{(q_0-\mu-|\bf{q}|)}{q_0^2-\bar{\epsilon}_q^2}+\frac{m}{2\mu}\frac{q_0+q_{||}}{q_0^2-\epsilon_q^2}+\gamma^0\frac{m^2}{2\mu}\biggl(\frac{q_0+q_{||}}{q_0^2-\epsilon_q^2}\biggr)^2\Lambda^-(q) \\
S_{12}(q)&=&-\frac{\mbox{\boldmath$M^{\dag}$}G^*(q)}{q_0^2-\epsilon_q^2}\Lambda^+(q)-\frac{G^*(q)}{2\mu(
q_0^2-\epsilon_q^2)}(\gamma^0\mbox{\boldmath$M^{\dag}$}m\Lambda^-(q)+\gamma^0m\mbox{\boldmath$M^{\dag}$}\Lambda^+(q))
\end{eqnarray}
$S_{22}$ and $S_{21}$ follow from $S_{11}$ and $S_{12}$ respectively
under the substitutions $\Lambda^\pm(q)\leftrightarrow
\Lambda^\mp(q),\pm\mu\leftrightarrow\mp\mu,\pm|{\bf q}|\leftrightarrow \mp
|{\bf q}|, \pm G(q)\leftrightarrow \mp G^*(q),
\mbox{\boldmath$M$}\leftrightarrow\mbox{\boldmath$M^{\dag}$} $. Note
that the antiparticle-gap does not appear to this order of the
expansion in $1/\mu$.
%\begin{center}
%{\bf Appendix B}
%\end{center}
\section{quark loop contribution}
The integral in Eq.~(\ref{qloop}) is most conveniently evaluated by contour
integration in the complex $q_0$ plane after noting that $d^4q=2\pi
q_{\perp}dq_{\perp}dq_{||}dq_0$ where $q_{\perp}=|{\bf q}_{\perp}|$;
${\bf q}_{\perp}={\bf q}-({\bf q.\hat{P}})\hat{\bf P}$ with ${\vec{\bf P}}$ being a vector
of magnitude $\mu$ directed normal to the Fermi surface. The
$dq_{\perp}$ integration runs from 0 to $\Lambda_{\perp}=2\mu$. The first
terms from $T_1$ and $T_2$ which correspond to the
particle-antiparticle piece then give 
\begin{equation}
{\rm Im}G_{\bar{q}q}(\omega)=\frac{6\mu^2}{\pi^2}\int_{-\infty}^{\infty}dq_{||}\delta(\omega-2\mu-\sqrt{q_{||}^2+\Delta^2})\biggl(1+\frac{q_{||}}{\sqrt{q_{||}^2+\Delta^2}}\biggl)~,
\end{equation}
where $dq_{||}>0$ pertains to particles while $dq_{||}<0$ to
holes. The $dq_{||}$ integration yields
\begin{equation}
{\rm Im}G_{\bar{q}q}(\omega)=12\frac{\mu^2}{\pi}\frac{1}{\sqrt{1-(\Delta/(\omega-2\mu))^2}}\Theta(\omega-(2\mu+\Delta))\Theta(\Lambda_*-\sqrt{(\omega-2\mu)^2-\Delta^2})~. \label{qqbar}
\end{equation} 
The branch cut at $\omega=2\mu+\Delta$ signifies the threshold for
particle-antiparticle excitations.\\
For the terms inside the square bracket of $T_1$ and $T_2$, we have
\begin{eqnarray}
G(Q_0)&=&-\frac{3(m_u^2+m_d^2)}{4\mu^2}\int\frac{d^4q}{(2\pi)^4}\biggl\{\biggl[\biggl(\frac{p_0+p_{||}}{p_0^2-\epsilon_p^2}\biggr)^2+(p_{||}\rightarrow
-p_{||}) \nonumber\\
&+&\frac{(k_0+k_{||})(p_0+p_{||})}{(k_0^2-\epsilon_k^2)(p_0^2-\epsilon_p^2)}+(k_{||}\rightarrow
-k_{||})\biggr]+(K\rightarrow P)\biggr\}~. \label{massins}
\end{eqnarray}
The arrangement of poles in the complex $q_0$ plane depends
on the external energy $\omega$. For the terms in the first
line of Eqn.(\ref{massins}), which arise from two mass insertions in
the same line, the residues from the poles exactly
cancel if $\omega>2\epsilon_q$. A finite residue is obtained in the
case $\omega<2\epsilon_q$ only. The interpretation is that a particle
disappears into the Dirac sea and reappears from it at a later time,
corresponding to the two mass insertions. The terms on the second line
which correspond to one mass insertion on each quark line, lack an
imaginary part and do not contribute to the spectral function. Note that we have used the large $\mu$ approximation
for antiparticles so that only particle lines can be cut (only
particle propagators appear in Eqn.(\ref{massins})). Moreover, the
external energy has to exceed the mass gap, so that the above
expression is valid only for $\omega$ typically greater than the $u$ and $d$
quark mass. Neglecting the energy dependence of the gap to a first
approximation (we will assume it to depend on the chemical potential
only), we find 
\begin{equation}
{\rm Im}G_m(\omega)=-\frac{3}{16\pi^2}(m_u^2+m_d^2)(w^2-\Delta^2)\int_{-\infty}^{\infty}\frac{dq_{||}}{\epsilon_q^3}\theta(2\epsilon_q-\omega)~.
\end{equation}
If $\omega<2\Delta , q_{||}$ is unrestricted whereas if
$\omega>2\Delta , |q_{||}|>\sqrt{\frac{\omega^2}{4}-\Delta^2}$ must be
satisfied (it should be noted that $q_{||}$ can be at least $-\mu$). Hence
\begin{equation}
{\rm Im}G_m(\omega)=\frac{3(m_u^2+m_d^2)}{8\pi^2}(w^2-\Delta^2)\biggl[\Theta(2\Delta-\omega)\frac{2}{\Delta^2}+\Theta(\omega-2\Delta)\biggl\{\biggl(\int_{-\mu}^{{\rm max}(-\mu,-\sqrt{\omega^2/4-\Delta^2})}+\int_{\sqrt{\omega^2/4-\Delta^2}}^{\Lambda_*}\biggr)\frac{dq_{||}}{\epsilon_q^3}\biggl\}\biggr]~.
\end{equation}
The third term from $T_1,T_2$ involving gap and mass insertions on the
same quark line evaluates to
\begin{equation}
\frac{(m_u^2+m_d^2+2m_s^2)\Delta^2}{2\pi^2}\biggl[\Theta(2\Delta-\omega)\frac{2}{\Delta^2}+\Theta(\omega-2\Delta)\biggl\{\biggl(\int_{-\mu}^{{\rm max}(-\mu,-\sqrt{\omega^2/4-\Delta^2})}+\int_{\sqrt{\omega^2/4-\Delta^2}}^{\Lambda_*}\biggr)\frac{dq_{||}}{\epsilon_q^3}\biggl\}\biggr]~. \label{massgapins}
\end{equation}
Finally, the diquark contribution reads
\begin{eqnarray}
\frac{m_um_d\Delta^2}{\pi^2}&\biggl[&\Theta(2\Delta-\omega)\frac{8}{\omega\sqrt{4\Delta^2-\omega^2}}{\rm
tan}^{-1}\biggl(\sqrt{\frac{4\Delta^2}{\omega^2}-1}\biggr) 
+\Theta(\omega-2\Delta)\biggl\{\frac{1}{\omega}\int_{{\rm
max}(-\mu,-\sqrt{\omega^2/4-\Delta^2})}^{\sqrt{\omega^2/4-\Delta^2}}\frac{dq_{||}}{(\omega^2/4-\epsilon_q^2)}
\nonumber \\
&+&\biggl(\int_{-\mu}^{{\rm max}(-\mu,-\sqrt{\omega^2/4-\Delta^2})}+\int_{\sqrt{\omega^2/4-\Delta^2}}^{\Lambda_*}\biggr)\frac{dq_{||}}{\epsilon_q(\epsilon_q^2-\omega^2/4)}\biggr\}\biggr]~. \label{gapins}
\end{eqnarray}
where $\omega$ is typically greater than the $u,d$ quark masses.  
%\begin{center}
%{\bf Appendix C}
%\end{center}
\section{$\sigma$ meson mass and width}
The relevant terms from the effective Lagrangian Eqn.(\ref{leff}) are
\begin{equation}
L_{eff}^{\bf\sigma}=A_1{\rm Tr}[(M+{\bf\sigma})U^\dag]{\rm
Tr}[(M+{\bf\sigma})U^\dag]+A_2{\rm
Tr}[(M+{\bf\sigma})U^\dag(M+{\bf\sigma})U^\dag];\quad U=\exp
(2i\pi^a\tau^a/f_\pi).
\end{equation}
Here, $M={\rm diag}(m_u,m_d,m_s)$ while the ${\bf\sigma}$ field is
decomposed as  ${\bf\sigma}=\sigma\otimes{\bf 1}_2^f$. The flavor
structure involves only the $u,d$ quarks. We expand the chiral field
$U$ to ${O}(\pi^2)$ and compare the coefficients of the $\sigma^2$
and $\sigma\pi\pi$ terms to the kinetic term and two-pion coupling
term of the $\sigma$ field.
\begin{equation}
\frac{1}{2}m_\sigma^2\sigma^2=\frac{3\Delta^2}{\pi^2}\sigma^2,\quad
g_{\sigma\pi\pi}i\epsilon^{ab3}\pi^a\pi^b\sigma=\frac{\delta
m}{f_\pi^2}\frac{6\Delta^2}{\pi^2}i\epsilon^{ab3}\pi^a\pi^b\sigma \quad,
\end{equation}  
from which the mass of the sigma and it's coupling to two pions follows(Eqn.(\ref{sigmaeff})).
The spectral function of the $\sigma$ meson can be modelled as a
Breit-Wigner resonance 
\begin{equation}
\rho_\sigma(\omega,{\bf\vec{q}}=0)=\frac{{\rm Im}\Sigma
(\omega)}{(\omega^2-m_\sigma^2-{\rm Re}\Sigma (\omega))^2+({\rm Im}\Sigma
(\omega))^2}
\end{equation} 
where $\Sigma (\omega)$ denotes the self-energy of the $\sigma$ meson
dressed by the pion loop. It's imaginary part relates to the decay
width $\Gamma_{\sigma\pi\pi}$ of the $\sigma$ meson which is determined by $g_{\sigma\pi\pi}$
\begin{equation}
\Gamma_{\sigma\pi\pi}=\frac{g_{\sigma\pi\pi}^2}{24\pi {m_\sigma}^2}\sqrt{\frac{m_\sigma^2}{4}-m_\pi^2}
\end{equation}
The real part of the self-energy is determined by the mass-shell
condition to be
\begin{eqnarray}
{\rm Re}
\Sigma_{\sigma\pi\pi}(\omega)&=&\frac{g_{\sigma\pi\pi}^2}{16\pi^2}\biggl[\sqrt{1-\frac{4m_\pi^2}{\omega^2}}{\rm
ln}\biggl|\frac{1+\sqrt{1-\frac{4m_\pi^2}{\omega^2}}}{1-\sqrt{1-\frac{4m_\pi^2}{\omega^2}}}\biggr|-2\biggl(\frac{p_0}{\omega_0}\biggr){\rm
ln}\biggl(\frac{\omega_0+p_0}{m_\pi}\biggr)\biggr]\nonumber \\
p_0&=&\sqrt{\frac{m_\sigma^2}{4}-m_\pi^2};\quad \omega_0=m_{\sigma}/2=\sqrt{m_\pi^2+p_0^2}
\end{eqnarray}
%\begin{center}
%{\bf Appendix D}
%\end{center}
\section{$\sigma$ meson contribution}
As the $\sigma$ meson couples to a quark-conjugate quark pair, a gap
insertion on a quark line is required for an overlap with a
quark-antiquark source. Moreover, a chirality flip via a mass
insertion is essential since left and right-handed quarks pair with
their own chirality species only. This necessitates retaining the
leading mass terms in the expansion of the propagators ($O(m^2)$ terms
need not be retained however). Since there is an overall factor of
$\Delta/f_\pi$ in the $\sigma$ coupling to the fermions, we expect the
final result to be proportional to the product of $\Delta^2/f_\pi$ and some
combination of the quark masses (including the strange quark mass
which comes in because of the color flavor locked structure of the
gap). Performing the trace over all internal symmetries, we obtain
\begin{eqnarray}
J_{\sigma\bar{\psi}\psi}&=&\frac{\Delta}{f_\pi}\int\frac{d^4q}{(2\pi)^4}\biggl\{\alpha\biggl[\frac{p_{||}\Delta(K)}{(k_0^2-\epsilon_k^2)(p_0^2-\epsilon_p^2)}+\frac{k_{||}\Delta(P)}{(k_0^2-\epsilon_k^2)(p_0^2-\epsilon_p^2)}\biggr]+\beta\biggl[\frac{\Delta(K)}{k_0^2-\epsilon_k^2}+\frac{\Delta(P)}{p_0^2-\epsilon_p^2}\biggr]\biggr\}\nonumber
\\
\alpha&=&\frac{4(m_u+m_d+2m_s)}{\mu};\quad \beta=\frac{4(m_u+m_d+2m_s)}{\mu^2} .
\end{eqnarray}
To simplify the momentum integration, the gap is assumed to be
energy-momentum independent, and can be pulled out of the
integration. As before, $d^4q=2\pi q_{\perp}dq_{\perp}dq_{||}dq_0$ and
the contour integration over complex $q_0$ is performed first. Then,
the terms proportional to $\alpha$ yield equal and opposite residues
in both cases $\omega>2\epsilon_q$ and $\omega<2\epsilon_q$. Thus, if
the mass and gap insertions are on different lines, the result is
zero. A finite coupling is obtained only when these two insertions
appear on the same quark line. in other words, only the mixed terms
($\sim m\Delta$) that appear in the expansion of the propagator
contribute. These are precisely the terms proportional to $\beta$,
which give
\begin{equation}
J_{\sigma\bar{\psi}\psi}=\frac{\beta\mu^2\Delta^2}{4\pi^2f_\pi}\int_{-\infty}^{\infty}\frac{dq_{||}}{\epsilon_q}\Theta(\epsilon_q-\frac{\omega}{2})\quad.
\end{equation}
For the case $\omega<2\Delta$, the $q_{||}$ integration is
unrestricted. Since typical momenta are of the order of the gap, we
may run the integration from $\Delta$ to $\Lambda_*$. For the case
$\omega>2\Delta, |q_{||}|>\sqrt{\frac{\omega^2}{4}-\Delta^2}$ must be
satisfied ($q_{||}$ can be at least $-\mu$). Hence
\begin{equation}
J_{\sigma\bar{\psi}\psi}=\frac{\Delta^2(m_u+m_d+2m_s)}{\pi^2f_\pi}\biggl[\Theta(2\Delta-\omega)2x_0+\Theta(\omega-2\Delta)
\biggl(\int_{-\mu}^{{\rm max}(-\mu,-\sqrt{\omega^2/4-\Delta^2})}+\int_{\sqrt{\omega^2/4-\Delta^2}}^{\Lambda_*}\biggr)\frac{dq_{||}}{\epsilon_q}\biggr]\quad,
\end{equation}
where $x_0={\rm ln}(\Lambda_*/\Delta)$.

%---------------------------------------------------------------------

\begin{flushleft}

%-----------------      FIGURES    -------------------
\end{flushleft}

\begin{thebibliography}{99}
\bibitem{Adler68} S.L. Adler and R. F. Dashen, {\it Current Algebras and
Applications to Particle Physics}, (W. A. Benjamin, New York, 1968).
\bibitem{Weinberg79} S. Weinberg, Physica (Amsterdam) {\bf 96A}, 327
(1979).
\bibitem{Gasser84} J. Gasser and H. Leutwyler, Ann. Phys. {\bf 158}, 142 (1984).
\bibitem{Yam96} H. Yamagishi and I. Zahed, Ann. Phys. {\bf 247}, 292 (1996).
\bibitem{ParticleDB2000} Review of Particle Physics, Eur. Phys. J.
{\bf C15}, 1 (2000).
\bibitem{GellMann60} M. Gell-Mann and M. Levy, Nuovo Cimento {\bf 16}, 705 (1960).
\bibitem{NJL61} Y. Nambu and G. Jona-Lasinio, Phys. Rev. {\bf 122} 345 (1961).
\bibitem{K01} T. Kunihiro, {\it Invited talk,
International Workshop on Physics with GeV electrons and gamma rays,
Laboratory of Nuclear Science, Tohoku University, Feb. 13-15} (2001).
\bibitem{Hatsuda99} T. Hatsuda, T. Kunihiro and H. Shimizu,
Phys. Rev. Lett. {\bf 82} 2840 (1999).
\bibitem{Schuck00} P. Schuck, Z. Aouissat, G. Chanfray and J. Wambach,
{\it International workshop XXVIII on Gross Properties of Nuclei and
Nuclear Excitations, Hirschegg, Austria} January 16-22 (2000).
\bibitem{Schuck88} P.Schuck, W. N\"orenberg, G. Chanfray,
Z. Phys. {\bf A330} 119 (1988).
\bibitem{Frau78}
S. C. Frautschi,
Asymptotic freedom and color superconductivity in dense quark matter,
in: Proceedings of the
Workshop on Hadronic Matter at Extreme Energy Density, N. Cabibbo, Editor,
Erice, Italy (1978).
\bibitem{Barrois77}
B.~C.~Barrois,
Nucl.\ Phys.\  {\bf B129}, 390 (1977).
\bibitem{Bailin84}
D.~Bailin and A.~Love,
Phys.\ Rept.\  {\bf 107}, 325 (1984).
\bibitem{Alford98}
M.~Alford, K.~Rajagopal and F.~Wilczek,
Phys.\ Lett.\  {\bf B422}, 247 (1998)
[hep-ph/9711395].
\bibitem{Rapp98}
R.~Rapp, T.~Sch\"afer, E.~V.~Shuryak and M.~Velkovsky,
Phys.\ Rev.\ Lett.\  {\bf 81}, 53 (1998)
[hep-ph/9711396].
\bibitem{alf} 
M.~G.~Alford, K.~Rajagopal and F.~Wilczek,
Nucl.\ Phys.\ B {\bf 537}, 443 (1999)
[hep-ph/9804403].
\bibitem{rapp} 
R.~Rapp, T.~Sch\"afer, E.~V.~Shuryak and M.~Velkovsky,
Annals of Phys.\  {\bf 280}, 35 (2000)
[hep-ph/9904353].
\bibitem{Schafer99}
T.~Sch\"afer,
Nucl.\ Phys.\ B {\bf 575}, 269 (2000)
[hep-ph/9909574].
\bibitem{Evans99}
N.~Evans, J.~Hormuzdiar, S.~D.~Hsu and M.~Schwetz,
Nucl.\ Phys.\ B {\bf 581}, 391 (2000)
[hep-ph/9910313].
\bibitem{Hong99}
D. K. Hong, M. Rho and I. Zahed, Phys. Lett. {\bf B468}, 261 (1999)
[hep-ph/9906551].
\bibitem{Casal99}
R.~Casalbuoni and R.~Gatto,
Phys.\ Lett.\ B {\bf 464}, 111 (1999)
[hep-ph/9908227].
\bibitem{Son} 
D.T. Son and M.A. Stephanov, Phys. Rev. {\bf D 61}, 074012 (2000)
[hep-ph/9910491].
Erratum: Phys. Rev. {\bf D}62 059902 (2000) [hep-ph/0004095].
[hep-ph/9910491].
\bibitem{Schafer}
T. Sch\"afer, Phys. Rev. D {\bf 65} 074006 (2002) [hep-ph/0109052].
\bibitem{Jai}
P. Jaikumar, R. Rapp and I. Zahed, Phys. Rev. C {\bf 65}, 055205 (2002).
[hep-ph/0112308].
\bibitem{Son99} 
D.~T.~Son and M.~A.~Stephanov,
Phys.\ Rev.\ D {\bf 61}, 074012 (2000).
\bibitem{Hong100}
D. K. Hong, Phys. Lett. B {\bf 473}, 118 (2000) [hep-ph/9812150]. 
\bibitem{Hong200}
D. K. Hong, Nucl. Phys. B {\bf 582}, 451 (2000) [hep-ph/9905523].
\bibitem{Gatto01}
R. Casalbuoni, R. Gatto and G. Nardulli, Phys. Lett. B {\bf 498}, 179
(2001) [hep-ph/0010321]; Erratum-ibid. B {\bf 517} 483 (2001).
\bibitem{Park99}
B.-Y. Park, M. Rho, A. Wirzba and I. Zahed, Phys. Rev. D{\bf 62},
034015 (2000)
[hep-ph/9910347].
\bibitem{Schulze}
D. Blaschke, Yu.L. Kalinovsky, S. Schmidt and H.-J. Schulze,
Phys. Rev. C {\bf 57} 438 (1998) [hep-ph/9709058].
\bibitem{Sannino01}
R. Casalbuoni, Z. Duan and F. Sannino, Phys. Rev. D {\bf 63} 114026 (2001) 
[hep-ph/0011394]. 
\bibitem{JPS02}
P. Jaikumar, M. Prakash and T. Sch\"afer, Phys Rev. D {\bf 66}, 063003
(2002) [hep-ph/0203088].
\bibitem{Pisarski99}
R. D. Pisarski and D. H. Rischke, Phys. Rev. D{\bf 60}, 094013 (1999)
[nucl-th/9903023].
%\bibitem{Hades}
%HADES, Technical Proposal, GSI 1994; \\
%J. Friese et al., Nucl. Phys. {\bf A654} 1017 (1999). 
\end{thebibliography}
\end{document}